\documentclass[
    prd, twocolumn, superscriptaddress, nofootinbib, amsmath, amssymb,
    aps
]{revtex4-2}

\usepackage{graphicx}
\usepackage{dcolumn}
\usepackage{bm}
\usepackage{physics}
\usepackage{color}
\usepackage{mathrsfs,amsfonts,hepunits, color}
\usepackage{xcolor}
\usepackage{braket}
\usepackage{MnSymbol}
\usepackage{hyperref}

\newcommand{\be}{\begin{equation}}
\newcommand{\ee}{\end{equation}}

\newcommand{\azero}{{\rm a}_0}

\begin{document}

\title{Measuring the molecular Migdal effect with neutron scattering on diatomic gases}

\author{Yonatan Kahn}
\email{yfkahn@illinois.edu}
\affiliation{Department of Physics, University of Illinois Urbana-Champaign, Urbana, IL 61801, U.S.A.}

\author{Jes\'{u}s P\'{e}rez-R\'{i}os}
\email{jesus.perezrios@stonybrook.edu}
\affiliation{Department of Physics and Astronomy, Stony Brook University, Stony Brook, NY  11794, U.S.A.}
\affiliation{Institute for Advanced Computational Science, Stony Brook University, Stony Brook, NY 11794, U.S.A.}

\date{\today}
\begin{abstract}

The Migdal effect is a key inelastic signal channel which could be used to detect low-mass dark matter, but it has never been observed experimentally using Standard Model probes. Here we propose a conceptual design for an experiment which could detect the Migdal effect in diatomic molecules through low-energy neutron scattering, and we provide the requirements on the beam spectrum and reducible backgrounds such that a detection may be achieved. The enhancement of the Migdal rate through non-adiabatic couplings, which are absent in isolated atoms, combined with the distinctive photon energies of electronic transitions in CO, suggest that a positive detection of the molecular Migdal effect may be possible with modest beam times at existing neutron facilities.

\end{abstract}
\maketitle

\section{Introduction}

The Migdal effect~\cite{migdal1941ionization}, where nuclear scattering induces an electronic excitation in an atom, molecule, or solid, has been studied theoretically for nearly a century but has never been conclusively observed experimentally. The main challenge is the very small rate compared to elastic scattering, combined with the difficulty of distinguishing a primary Migdal event from a secondary electronic excitation or ionization following ordinary elastic nuclear scattering. The Migdal effect has been proposed to search for sub-GeV dark matter as a way to evade nuclear recoil thresholds via an electronic excitation signal~\cite{Bernabei:2007jz,Ibe:2017yqa,Dolan:2017xbu,Bell:2019egg,Baxter:2019pnz,Essig:2019xkx,Liang:2019nnx,GrillidiCortona:2020owp,Liu:2020pat,Knapen:2020aky,Liang:2020ryg,Wang:2021oha,Liang:2022xbu,Berghaus:2022pbu,Essig:2022dfa}, but first this effect must be observed with Standard Model probes in order to calibrate it~\cite{Bell:2021ihi,Araujo:2022wjh,Adams:2022zvg,Bell:2023uvf,Xu:2023wev}.

In this paper, motivated by recent developments on the \emph{molecular} Migdal effect relevant for dark matter detection~\cite{Blanco:2022pkt}, we propose a new concept for measuring the Migdal effect. A low-energy ($\sim 100 \ {\rm eV}$) neutron beam is used to induce \emph{bound} Migdal transitions through nuclear scattering in a molecular gas such as carbon monoxide (CO), with a probability of $\sim 10^{-5}$ per neutron scattering event, leading to the emission of UV and visible photons with a distinctive spectrum that can be efficiently detected. As we will show, such transitions are extremely rare if induced by secondary collisions following a primary elastic scattering event, and the transitions we focus on are well-separated from nearby molecular transitions such that the only irreducible source of background is absorption of $\sim 10$ eV photons to directly drive the transition. By using a beam such as the Spallation Neutron Source (SNS) at Oak Ridge National Laboratory with a definite time structure, time correlations can be exploited to further reduce backgrounds not associated with the source. We emphasize the our goal in this paper is not to propose a concrete experimental setup, but rather (in a similar spirit to Ref.~\cite{lovesey1982electron}) to perform a conceptual study in order to determine the requirements on such a setup which could render the Migdal effect observable.

   \begin{figure}
        \centering
        \includegraphics[width=1\linewidth]{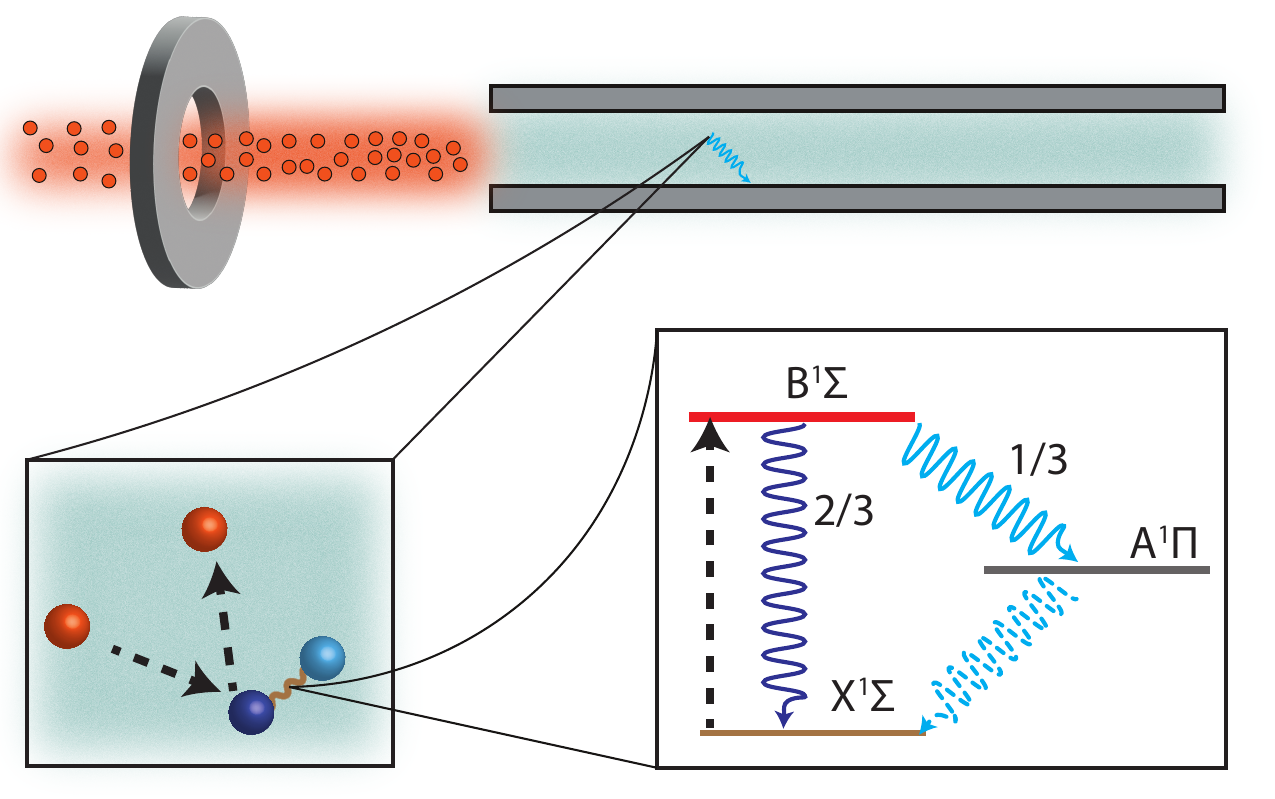}
        \caption{Sketch of the experimental proposal. A collimated neutron beam collides with a gas of CO molecules. A neutron-molecule collision via the molecular Migdal effect induces the X--B transition in CO. This excitation has two decay pathways: one going back to the X state and another ending up in the A state. The Migdal signal is the B--A transition.}
        \label{fig1}
    \end{figure}

\section{Schematic of proposed experiment}
Our experimental concept is schematically represented in Fig.~\ref{fig1}. A low-energy, highly-collimated neutron beam traverses a $\sim 1$ m long cell of cross-sectional area $\sim 1 \ {\rm cm}^2$ filled with carbon monoxide (CO) molecules in the gas phase. Following a Migdal scattering event, the molecule undergoes the transition X$^1\Sigma \to $B$^1\Sigma$ with an electronic energy gap of $10.78 \ {\rm eV}$, whose decay leads to photon emission following two pathways. Pathway (1) (dark blue), the de-excitation back to the ground electronic state B$^1\Sigma \rightarrow $X$^1\Sigma$, has a branching ratio of $\approx 2/3$ and can induce the inverse process in nearby CO molecules, leading to a loss of signal from attenuation. On the other hand, pathway (2) (light blue), with a branching ratio of $\approx 1/3$, results in visible or near-UV photon emission peaking at $\sim 400 \ {\rm nm}$ from B$^1\Sigma \to $A$^1\Pi$ which streams freely, while the subsequent 158 nm photon from  A$^1\Pi \to $X$^1\Sigma$ can excite the inverse transition in adjacent CO molecules and is attenuated. Therefore, our signal is the B$^1\Sigma \rightarrow $A$^1\Pi$ transition from (2), shown in solid light blue in Fig.~\ref{fig1}. The signal photons can be detected with large-area photomultiplier tubes (PMTs) lining the walls of the gas cell, which have high efficiency at the signal wavelength.

The pressure $p$ and temperature $T$ of the gas cell should be chosen such that black-body radiation effects are irrelevant, collision-induced absorption is highly suppressed, and no collisional quenching of electronic transitions occurs. For simplicity, we assume the gas cell is at room temperature ($T=300$~K), but in a realistic implementation the temperature will be set by the requirements of the photodetector. The B$^1\Sigma$ state of CO has a lifetime $\tau_B = 22.3\pm 0.5$ ns, which is typical of allowed molecular transitions. However, there is another zero-photon decay channel that can de-excite the molecule: CO-CO collisions, also known as self-quenching collisions. In the case of CO, its self-quenching cross section for the B state has been calculated to be $\sigma_{\text{self}}=132~{\rm a}_0^2$ \cite{COCO}. The typical collision time is $t_{\text{coll}}=\frac{1}{n \langle \sigma_{\text{self}} \rangle \langle v_{\rm CO} \rangle}$, where $\langle v_{\rm CO} \rangle$ denotes the average thermal velocity of CO molecules, $n$ is the gas number density and $\langle \sigma_{\text{self}} \rangle$ is the thermally-averaged self-quenching cross section, which in this case we approximate as the constant $\sigma_{\text{self}}$. At $T=300$~K, imposing that $t_{\text{coll}} > \tau_B$, we find that $n_{\rm CO} < 3\times 10^{16}$~cm$^{-3}$, which sets an upper bound on the pressure at $p \sim 1$~mbar.

\section{Estimated rate} 
Calculating the expected rate of Migdal excitation is a straightforward modification of the formalism of Ref.~\cite{Blanco:2022pkt}, which computed dark matter-induced Migdal excitation in heteronuclear diatomic molecules, including to the same B$^1\Sigma$ state of CO. Throughout the rest of the analysis, we use particle physics units with $\hbar = c = 1$. The probability of Migdal excitation to an electronic state $\alpha$ with energy $\epsilon_\alpha$ through nuclear scattering, via the non-adiabatic coupling channel, decomposes into electronic and nuclear matrix elements, $P_M^{(\alpha)}(\vec{q}) = P^{(\alpha)}_{e}(\vec{q}) \times P^{(\alpha)}_{N}(\vec{q})$, which depend on the momentum transfer $\vec{q}$. As was demonstrated in Ref.~\cite{Blanco:2022pkt}, the non-adiabatic coupling channel dominates the Migdal rate in CO compared to other transitions by two orders of magnitude. At 300 K, most of the molecules will be in the vibrational ground state; for simplicity, we will neglect rotational excitations which amounts to assuming that the initial and final rotational states of the molecule are the same, yielding a lower bound on the total rate. The electronic matrix element is
\be
\label{eq:PeNAC}
P^{(\alpha)}_{e} = \frac{q^2 \eta^2 |G_{\alpha 0}|^2}{M^2(\epsilon_\alpha - \epsilon_0)^2},
\ee
where $q = |\vec{q}|$, $\eta$ is the cosine of the angle between $\vec{q}$ and the molecular axis, $M$ is the total nuclear mass of the molecule, $\epsilon_0$ is the ground-state energy, and $G_{\alpha 0}$ is the non-adiabatic coupling between $\alpha$ and the ground state, equal to $1.50 \ \azero^{-1}$ for the X--B transition in CO. The nuclear matrix element is~\cite{colognesi2005can,Blanco:2022pkt}
\begin{align}
\label{eq:PNNAC}
P^{(\alpha)}_{N} = \sum_{n= 0}^{n_{\rm diss.}} &\Bigg( \bigg |\bigg\langle \chi^{(\alpha)}_n \bigg | a_1 \frac{M_2}{\mu} e^{- i \frac{\mu }{M_1} q \rho \eta }  \nonumber \\  
&\;\;\;\;-a_2 \frac{M_1}{\mu} e^{+ i \frac{\mu }{M_2} q \rho \eta } \bigg | \chi_0 \bigg\rangle \bigg |^2\Bigg),
\end{align}
where $\chi_0$ is the nuclear ground state wavefunction, $\chi_n^{(\alpha)}$ are the vibrational wavefunctions at harmonic oscillator level $n$ associated to state $\alpha$ in the Born-Oppenheimer approximation, $n_{\rm diss.}$ is the level at which the vibrational energy exceeds the dissociation energy of the molecule, $M_1$ and $M_2$ are the masses of the two nuclei, $\mu = M_1 M_2/M$ is the nuclear reduced mass, $\rho$ is the internuclear separation, and $a_1$ and $a_2$ are dimensionless coupling constants characterizing the interactions between the scattering probe and the nuclear targets. In the case of neutron scattering, it is more convenient to work in terms of the (dimensionful) neutron scattering lengths off free targets~\cite{sears1992neutron}, $b_{\rm C} = 6.64 \ {\rm fm} \times \frac{12}{13}$ and $b_{\rm O} = 5.80 \ {\rm fm} \times \frac{16}{17}$, where the fractional factor accounts for the reduced mass between the neutron and the nucleus. Replacing $a_1$ and $a_2$ with scattering lengths $b_1$ and $b_2$ gives the nuclear matrix elements the dimension of a cross section, so instead of computing Migdal excitation probabilities, we deal directly with the neutron-induced Migdal cross section:
\begin{align}
& \sigma^{(\alpha)}_M(\vec{q}) =  \frac{q^2 \eta^2 |G_{\alpha 0}|^2}{M^2(\epsilon_\alpha - \epsilon_0)^2} \times \nonumber \\
& \sum_{n= 0}^{n_{\rm diss.}} \Bigg( \bigg |\bigg\langle \chi^{(\alpha)}_n \bigg | b_1 \frac{M_2}{\mu} e^{- i \frac{\mu }{M_1} q \rho \eta } -b_2 \frac{M_1}{\mu} e^{+ i \frac{\mu }{M_2} q \rho \eta } \bigg | \chi_0 \bigg\rangle \bigg |^2\Bigg).
\label{eq:sigmaM}
\end{align}

To obtain the total scattering rate, we integrate over the neutron spectrum and the possible momentum transfers. In Ref.~\cite{Blanco:2022pkt}, the molecules were taken to have a fixed orientation and the dark matter spectrum was paramterized by a velocity distribution. To make contact with this formalism, we can convert the neutron energy spectrum to a velocity distribution, which we pretend is isotropic in direction in order to model the random orientations of the target molecules. As a model for low-energy neutrons, we assume an $E^{-1}$ spectrum between energies $E_{\rm min}$ and $E_{\rm max}$~\cite{10.1063/1.4953612,iverson2003spallation}, leading to an effective neutron velocity distribution of
\be
f_n(\vec{v}) = \frac{2}{\ln(E_{\rm max}/E_{\rm min})} \frac{1}{4\pi v^3} \Theta(v - v_{\rm min}) \Theta(v_{\rm max} -v),
\label{eq:fn}
\ee
where $v_{\rm min, max} = \sqrt{2 E_{\rm min, max}/m_n}$. The first prefactor in Eq.~(\ref{eq:fn}) ensures that $\int d^3 v \, f(\vec{v}) = 1$.  In terms of this velocity distribution, the total rate per target molecule is given by Fermi's Golden Rule as
\be
R= \frac{2\pi \rho_n}{m_n^3}\int d^3 v f(\vec{v}) \int \frac{d^3 q}{(2\pi)^3} \sigma^{(\alpha)}_M(\vec{q}) \delta(\Delta E - \omega_{\vec{q}}),
\ee
where $\rho_n$ is the mass density of neutrons in the target volume, $\omega_{\vec{q}} = \vec{q} \cdot \vec{v} - \frac{q^2}{2m_n}$ accounts for the non-relativistic kinematics of the scattering, and $\Delta E$ is the total excitation energy, equal to $\epsilon_\alpha - \epsilon_0$ plus the vibrational excitation energy. Strictly speaking, there should be a different $\Delta E_n$ for each vibrational level, so the energy-conserving delta function gets pulled inside the sum which defines $\sigma_M$. The final modification to the dark matter formalism arises because the neutron flux $\Phi_n = \rho_n v/m_n$, rather than $\rho_n$, is the experimentally-provided quantity. The factor of $v$ gets pulled inside the velocity integral, yielding our final expression for the rate per target molecule, 
\be
\label{eq:RFinal}
R= \frac{\Phi_n}{m_n^2}\int d^3 v \, \frac{f(\vec{v})}{v} \int \frac{d^3 q}{4\pi^2} \, \sigma^{(\alpha)}_M(\vec{q}) \delta(\Delta E - \omega_{\vec{q}}).
\ee

To evaluate Eq.~(\ref{eq:RFinal}), we need to compute matrix elements in Eq.~(\ref{eq:sigmaM}) with respect to various vibrational states. The true vibrational wavefunctions are well-approximated by the Morse potential, but analytic formulas which facilitate numerical calculations are available if we approximate the vibrational states by the best-fitting harmonic oscillator states. We follow the prescription of Ref.~\cite{Blanco:2022pkt}, and fit the ground state $\chi_0$ and the excited vibrational states $\chi^{(\alpha)}_n$ for $0 \leq n \leq 10$ to harmonic oscillator states at the same level $n$, letting us express the nuclear matrix elements analytically in terms of Hermite polynomials.\footnote{As described in more detail in Ref.~\cite{Blanco:2022pkt}, the choice of nuclear wavefunction has a $\mathcal{O}(1)$ effect on the total rate and thus the harmonic oscillator approximation should suffice for an order-of-magnitude estimate.} Using the energy eigenvalues of the Morse potential, there are in fact $n_{\rm diss.} = 69$ vibrational levels available before exceeding the dissociation energy of 14 eV, so summing only up to $n = 10$ likely underestimates the rate slightly. Taking $E_{\rm min} = 1 \ {\rm eV}$, $E_{\rm max} = 100 \ {\rm eV}$, and $\Phi_n = 5 \times 10^{8}/{\rm cm}^2/{\rm s}$ (a typical neutron flux from sources such as the SPS~\cite{10.1063/1.4953612}), and imagining a sample container with cross-sectional area $A = 1 \ {\rm cm}^2$ and length $L = 1 \ {\rm m}$, which contains $N_T = 2.3 \times 10^{18}$ molecular targets at the maximum density of $2.3 \times 10^{16}/{\rm cm}^3$, we find a total rate
\be
R_{\rm total} \approx 2.6 \ {\rm events/min}
\ee
after accounting for the $\approx 1/3$ branching ratio of the signal pathway. If the neutron spectrum is chopped at 18 eV to reduce high-energy scattering events that could lead to ionization or dissociation (discussed further below), the Migdal rate is reduced by a factor of 10. The total neutron elastic scattering rate for this target density and volume is approximately $R_{\rm el.} \approx \Phi_n (4\pi b_n^2) N_T = 5.2 \times 10^3 \ {\rm events/s}$, where $b_n \approx 6 \ {\rm fm}$ is an approximate average scattering length, so the Migdal rate is five orders of magnitude smaller. Nonetheless, the distinctive photon spectrum of the Migdal de-excitation may help to reduce backgrounds, as we will discuss below.

\begin{figure}[t]
        \centering        \includegraphics[width=0.5\textwidth]{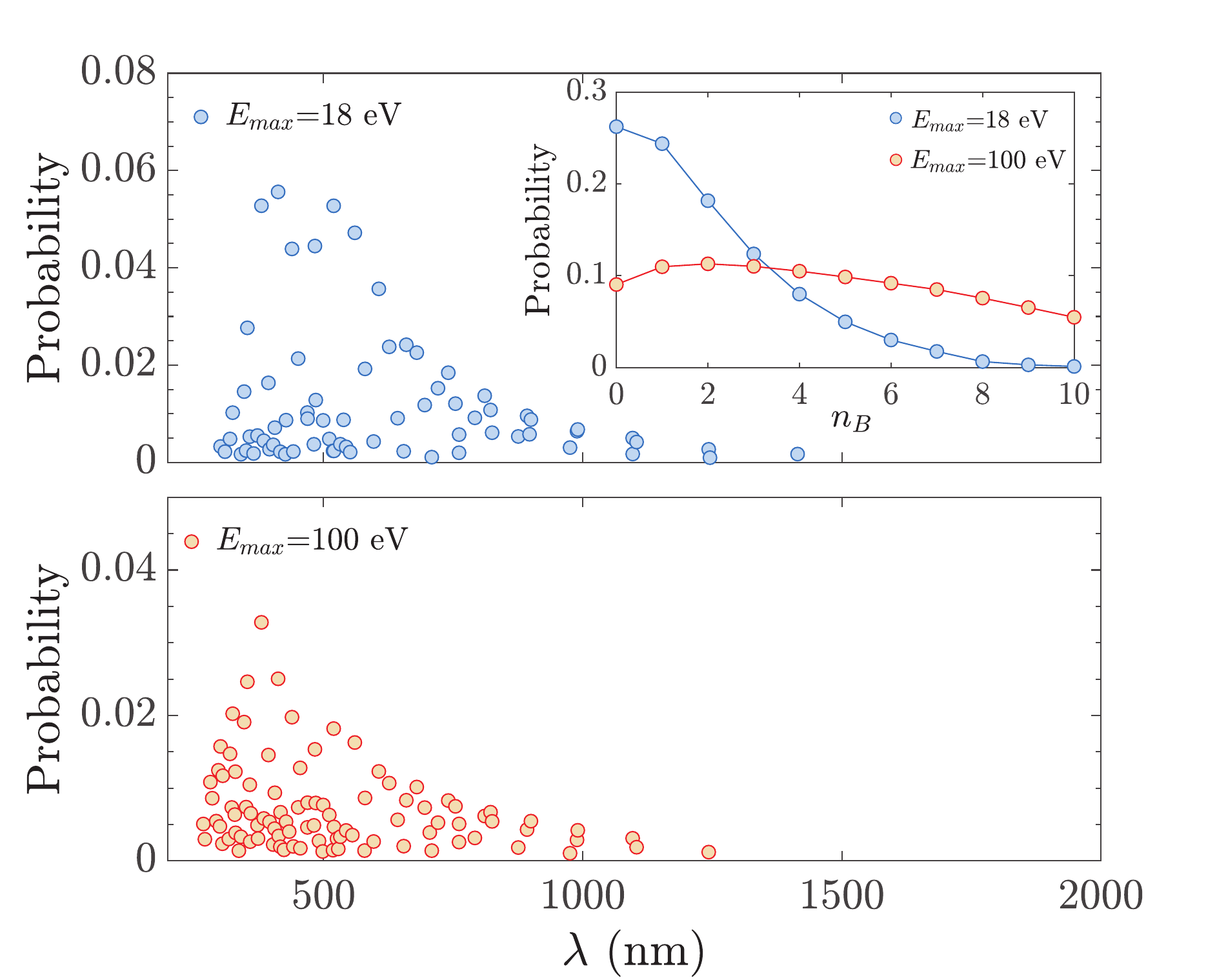}
        \caption{The spectrum of signal photons resulting from the $B \to A$ transition after Migdal excitation of CO molecules. The top panel (blue) assumes $E_{\rm max}  = 18 \ {\rm eV}$, while the bottom panel (red) assumes $E_{\rm max} = 100 \ {\rm eV}$. The inset shows the probability of $0_X \to n_B$ transitions, responsible for the shape of the spectrum. The emission spectrum is characterized by a series of extremely narrow lines corresponding to vibrational splittings of the $B$ and $A$ levels, of which a representative subset of transitions is shown. The width of each emission line is much smaller than the marker shown on the plot, such that the spectrum is effectively a sum of delta functions at each marked point.}
        \label{fig_wavlengths}
        \vspace{-0.2cm}
    \end{figure}

\section{Signal photon spectrum.} 
At $T=300$~K, most molecules are in the vibrational ground state $n_X=0$. Using Eq.~(\ref{eq:RFinal}), we can calculate the relative probabilities of transitions from $0_X$ to each $B$ vibrational state $n_B$, accounting for the expected neutron velocity distribution and integrating over allowed momentum transfers. The results are shown in the inset of Fig.~\ref{fig_wavlengths} for two different choices of neutron energy cutoff. For clarity, we only plot the de-excitation probabilities for a representative subset of transitions; the widths of these transitions are extremely narrow, so the spectrum appears as a series of narrow lines rather than a continuum. If the neutron spectrum extends to 100 eV (red), the final vibrational states are roughly equipartitioned, while if the neutron spectrum is chopped at 18 eV (blue), the transitions tend to end up in the ground vibrational state $0_B$. The vibrational energies of the $B$ states with respect to the $0_X$ ground-state energy range from 0.13 eV for $0_B$ to $2.5 \ {\rm eV}$ for $n_B = 10$.

Next, a given $n_B$ state will decay to the A$^1\Pi$ state following the usual dipole selection rules. The emission intensity depends on the Franck-Condon factor, which allows us to calculate the spectral lineshape of the Migdal signal. The results, shown in the top panel of Fig.~\ref{fig_wavlengths} for $E_{\rm max} = 18 \ {\rm eV}$, show a distinctive photon spectrum centered at visible photon wavelengths that the PMTs can easily detect, with the spectrum shifted slightly toward the near UV for the higher-energy neutrons with $E_{\rm max} = 100 \ {\rm eV}$ (bottom panel). One can imagine taking advantage of the spectral properties of the Migdal signal in optical wavelengths by using various color filters on the PMTs, allowing the signal to be decoupled from the PMT dark rate to some extent, which would not be expected to exhibit a spectrum correlated with the Migdal spectrum.

   \begin{figure}[t!]
        \centering
        \includegraphics[width=1\linewidth]{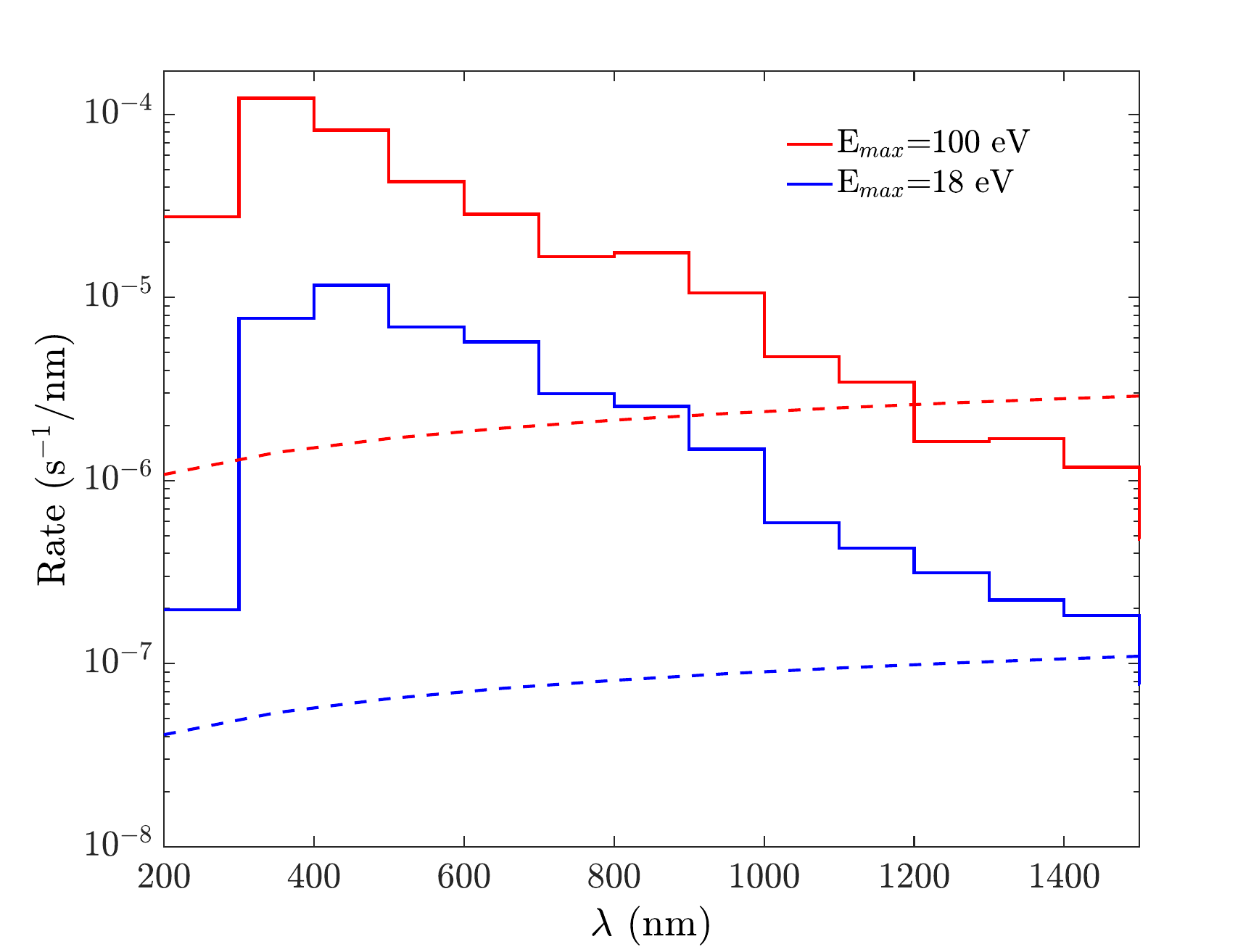}
        \caption{Rate per unit wavelength of collision-induced emission, for $E_{\text{max}}=100$ eV (red dashed) and $E_{\text{max}}=18$ eV (blue dashed). The Migdal signal rate (solid lines, binned in 100 nm bins for visual clarity) is at least an order of magnitude larger for all wavelengths detectable by PMTs.}
        \label{fig3}
    \end{figure}

\section{Backgrounds}
The dominant background for our signal process will depend sensitively on the characteristics of the neutron beam, which we do not attempt to model here. Rather, we will argue that certain backgrounds are almost certainly negligible, while the presence of others imposes requirements on the associated photon spectrum of the beam.

First, photon absorption at $\sim 11 \ {\rm eV}$ can directly drive the X$^1\Sigma \to $B$^1\Sigma$ transition. Based on the oscillator strength for the 0-0 transition~\cite{Kirby1989}, in order for this rate to be below the signal rate, the photon flux at these energies would need to be $\lesssim 10^{-8} \, \Phi_n$. Even if the flux of photons generated along with beam were larger than this factor, it should be possible to attenuate the UV photons significantly with a thin layer of lead placed in front of the gas cell; the photon attenuation length at 10 eV is on the order of nm, while the neutron mean free path is many orders of magnitude larger. By doing a preliminary measurement with the gas cell emptied of target molecules, it may be possible to estimate the flux of these UV photons using the same PMTs that are used to detect the signal photon, thus giving an in-situ estimate of the background rate. We note that the large energy of this transition makes thermal photons a negligible source of background, since at 300 K, there is a Boltzmann suppression of $\sim e^{E_\gamma/T}=e^{-400}$. The first electronic state with the same spin as the ground state is the A$^1\Pi$ state, at 8.068~eV above the minimum of the ground electronic state~\cite{Database,Herzberg}, with a transition wavelength of 158 nm. Therefore, there is no direct electronic transition from the ground state in the same wavelength range as the Migdal signal. Vibrational transitions may occur, but these are highly suppressed due to a negligible Franck-Condon factor and very small dipole moment. We expect that if the UV photon flux is sufficiently small to not drive the X$^1\Sigma \to $B$^1\Sigma$ transition at a rate above the neutron-induced Migdal rate, any other such transitions will occur at a lower rate.

Higher-energy X-ray photons, which are generically expected to accompany neutron beams, may also drive the same transition through Compton scattering. While a full study of these backgrounds is beyond the scope of this work, we estimate here the requirements on such a background for this proposal to be viable. The Compton cross section for X-rays on free electrons is $\mathcal{O}(\rm barn)$~\cite{Comptonatoms}, which is comparable to the elastic neutron cross section. However, not every Compton scattering event will lead to the signal transition in CO; in noble atoms, bound-bound transitions from Compton scattering are typically 0.1\% to 1\% of the free-electron Compton cross section~\cite{amusia2002cross}. Therefore, to avoid this source of backgrounds with a Migdal signal rate that is $10^{-5}$ of the elastic scattering rate, one would need to ensure that the beam-correlated X-ray photon flux is $\lesssim 10^{-3} \Phi_n$. Extremely short-lived isotopes and sufficiently long-lived isotopes resulting from the neutron beam activating the material surrounding the gas target will also contribute an X-ray background that appears in the beam window. A typical value for $\sim$keV energy events observed by dark matter detection experiments near the surface is $\mathcal{O}(10^5)$ counts/kg/day/keV~\cite{SENSEI:2020dpa}. Scaling this rate up by a factor $A$ to account for beam activation of the surrounding material, and integrating this spectrum up to energies of $m_e = 511$ keV (at which point the Compton cross section begins to fall off with photon energy), we find a flux of $A \times 6.5 \times 10^{-5}/{\rm s}$ for our target mass of $1.1 \times 10^{-7} \ {\rm kg}$. Thus, we can tolerate an activation factor $A < 10^{10}$ to ensure that the beam-uncorrelated X-ray flux is below $10^{-3} \Phi_n$. We conclude that beam-correlated X-ray backgrounds are likely the dominant concern.

After an elastic neutron-CO collision, the CO molecule could subsequently collide with another CO molecule, leading to collision-induced emission in the same wavelength as our signal. The emission rate of photons with frequency $\omega$ is given by

\begin{equation}
A(\omega)=\frac{4\omega^3}{3}n_{\text{CO}}\int_{a}^{\infty}b\, db  \Big \langle v \, |\mathbf{d}(v,b,\omega)|^2  \Big \rangle ,
\end{equation}
where $v$ is the relative velocity of two CO molecules, $b$ is the impact parameter, $a$ is the diameter of a CO molecule, and $\mathbf{d}(v,b,\omega)$ is the Fourier transform of the transition dipole moment between the ground and the electronic state under consideration, and $\langle. \rangle$ denotes an average over the relative velocity distribution. As a first estimate, we assume that the transition dipole moment is independent of the impact parameter, leading to the geometric cross section

\begin{equation}
\label{eq9}
    A(\omega)=\frac{2a^2\omega^3}{3}n_{\text{CO}}\Big \langle v \, |\mathbf{d}(v, \omega)|^2  \Big \rangle. 
\end{equation}
Next, following the work of Karman et al.~\cite{Karman2018}, it is possible to find an analytical expression for $|\mathbf{d}(v,\omega)|^2$, assuming that the excitation is due to short-ranged exchange interactions, yielding 

\begin{widetext}
\begin{equation}
\label{eq10}
|\mathbf{d}(v,\omega)|^2= \biggr\rvert \frac{4 \mu d_0 e^{-\gamma a}\gamma\sqrt{\mu v}(\mu^2 v^2+2\mu \omega)^{1/4}}{\left( \gamma^2+\mu^2 v^2\right)^2+2( \gamma^2-\mu^2 v^2 )(2\mu \omega+\mu^2 v^2)+(\mu^2 v^2+2 \mu \omega)^2}\biggr\rvert^2,
\end{equation}
\end{widetext}
 where $\mu$ is the reduced mass of the CO-CO system. In Eq.~(\ref{eq10}), the transition dipole moment is assumed to scale as $d=d_0e^{-\gamma r}$, where $d_0$ is the transition dipole moment in the asymptotic region and $\gamma$ is the length scale of the dipole surface, which we estimate as $\gamma=3$ a$_0^{-1}$ and $d_0=1$ Debye. Finally, we calculate the speed distribution of the scattered CO molecules $f_{\rm CO}(v)$ using the kinematics of $2 \to 2$ hard-sphere scattering; this is formally identical to the standard calculation for dark matter-induced nuclear recoil~\cite{Lewin:1995rx}. The total rate, assuming every CO molecule undergoes a primary neutron scattering, is

\begin{equation}
\label{eq9}
    A(\omega)=\frac{2a^2\omega^3}{3}n_{\text{CO}}\int f_{\rm CO}(v) \, |\mathbf{d}(v,\omega)|^2 \, dv.
\end{equation}
The emission spectra for our two choices of maximum neutron beam energy are shown in Fig.~\ref{fig3}, along with the signal spectrum binned in 100 nm bins for visual clarity. Integrating the spectra over the range 250 -- 800 nm, which contains the majority of the Migdal signal and represents a reasonable window for PMT detection, yields background rates of $9.2 \times 10^{-4}/{\rm s}$ for $E_{\rm max} = 100 \ {\rm eV}$ and $1.5 \times 10^{-5}/{\rm s}$ for $E_{\rm max} = 18 \ {\rm eV}$. Compared to the expected Migdal rates of $0.13/{\rm s}$ for $E_{\rm max} = 100 \ {\rm eV}$ and $0.013/{\rm s}$ for $E_{\rm max} = 18 \ {\rm eV}$, the photon emission rate from secondary collisions is suppressed by more than two orders of magnitude in both cases, and thus is likely negligible in our setup.

Finally, we note that assuming the aforementioned backgrounds can be mitigated, the dark rate of the PMTs will determine the required data-taking time for a given detection significance. As an example of the dark rate characteristic of current state-of-the-art PMTs, the Hamamatsu 8-inch R14688-100 PMT has a dark rate of about 2 kHz at room temperature~\cite{Kaptanoglu:2023ayz}. Assuming that the dark counts are uncorrelated Poisson noise, and that the PMT dark rate may be measured with the neutron beam off for a sufficient time to reduce the systematic uncertainty below the statistical fluctuations, achieving a detection sensitivity of $S/\sqrt{B} = 3$ for a neutron beam with $E_{\rm max} = 100 \ {\rm eV}$ would require about 80 hours of beam time. This could be further reduced by cooling the PMTs below room temperature to lower the dark rate (though this would come at the cost of additional experimental infrastructure), or by using optical filters to statistically distinguish the Migdal spectrum peaking at 400 nm from a spectrally-flat background of dark counts.

\section{Conclusions}
In this paper we have laid out a detection strategy for the molecular Migdal effect using low-energy neutron beams scattering off diatomic CO gas. The same approach is undoubtedly applicable to other diatomic gases, such as N$_2$. However, further data on electronic quenching cross sections are required to make realistic predictions. Similarly, it could be possible to use highly polar molecules and, via a static electric field, exploit the distinctive directionality of the Migdal effect as studied in Ref.~\cite{Blanco:2022pkt}. This setup could also further reduce backgrounds and avoid the reabsorption of photons coming from the pathway (1), although the implementation is necessarily more complex than the present proposal. One could also consider using a buffer gas, such as He, to broaden the electronic transitions~\cite{Essig:2019kfe}, allowing for larger densities of molecules and thus a higher Migdal signal rate. We look forward to detailed follow-up studies on the feasibility of our proposed setup at neutron facilities such as the SNS.

{\it Acknowledgments.} We thank Dan Baxter for invaluable discussions throughout the completion of this work. The work of Y.K. was supported in part by DOE grant DE-SC0015655. J.P.-R. acknowledges the support of the Simons Foundation.

\bibliography{Migdal}

\end{document}